\documentclass[twocolumn,showpacs,preprintnumbers,amsmath,amssymb]{revtex4}

\usepackage{graphicx}
\usepackage{dcolumn}
\usepackage{bm}

\begin{document}

\preprint{}

\title{Optical Control of Field-Emission Sites by Femtosecond Laser Pulses}

\author{Hirofumi Yanagisawa${}^{1}$}
\email{hirofumi@physik.uzh.ch}
\author{Christian Hafner${}^{2}$}
\author{Patrick Don\'{a}${}^{1}$}
\author{Martin Kl\"{o}ckner${}^{1}$}
\author{Dominik Leuenberger${}^{1}$}
\author{Thomas Greber${}^{1}$}
\author{Matthias Hengsberger${}^{1}$}
\author{J\"{u}rg Osterwalder${}^{1}$}
\affiliation{${}^1$Physik Institut, Universit\"{a}t Z\"{u}rich, Winterthurerstrasse 190, CH-8057 Z\"{u}rich, Switzerland \\
${}^2$Laboratory for Electromagnetic Fields and Microwave Electronics, ETH Z\"{u}rich, Gloriastrasse 35, CH-8092 Z\"{u}rich, Swizerland}

\begin{abstract}
We have investigated field emission patterns from a clean tungsten tip apex induced by femtosecond laser pulses. Strongly asymmetric modulations of the field emission intensity distributions are observed depending on the polarization of the light and the laser incidence direction relative to the azimuthal orientation of tip apex. In effect, we have realized an ultrafast pulsed field-emission source with site selectivity on the 10 nm scale. Simulations of local fields on the tip apex and of electron emission patterns based on photo-excited nonequilibrium electron distributions explain our observations quantitatively. 
\end{abstract}

\pacs{79.70.+q, 73.20.Mf, 78.47.J-, 78.67.Bf}
\date{\today}
\maketitle

%%%%%%%%%%%%%%%%%%%%%%%%%%%%%%%%%%%%%%%%%%%%%%%%%%%%%%%%%%%%%%%%%%%%%%%%%%%%%%%%%%%%%%%%%%%% INTRODUCTION %%%%%%%%%%%%%%%%%%%%%%%%%%%%%%%%%%%%%%%%%%%%%%%%%%%%%%%%%%%%%%%%%%%%%%%%%%%%%%%%%%%%%%%%%%%%%%%%%%%%%%%%%%
Applying strong electric fields to a metal enables field emission due to electron tunneling into the vacuum. Field emission from metallic tips with nanometer sharpness is particularly interesting due to the high brightness and coherence of the electron beams \cite{gomer93, fursey03, fink86, fu01,cho04, oshima02}. When a focused laser pulse illuminates the tip, optical electric fields are enhanced at the tip apex (\emph{local field enhancement}) due to the excitation of surface electromagnetic (EM) waves like, e.g., surface plasmon polaritons. Only recently, it was found that the enhanced fields induce pulsed field emission in combination with a moderate DC voltage applied to the tip \cite{hommelhoff06a, hommelhoff06b, ropers07, barwick07}. Depending on the strength of the enhaced field, different field emission processes become dominant \cite{hommelhoff06a}. For relatively weak fields, single electron excitations by single- or multiphoton absorption are prevalent, and photo-excited electrons are tunneling through the surface potential barrier (\emph{photo-field emission}). On the other hand, very strong fields largely modify the tunneling barrier and prompt the field emission directly (\emph{optical field emission}).

So far, the different emission processes were disputed in the literature, while the local field enhancement was treated as a static effect of the laser field such as the lightening rod effect \cite{hommelhoff06a,hommelhoff06b, novotny02, hecht05}. However, dynamical effects are predicted to occur when the tip size is larger than approximately a quarter wavelength \cite{martin01}. Here we used a tip whose apex was approximately a quarter wavelength, and found that dynamical effects substantially influence the symmetries of field emission intensity distributions. At the same time, we have realized an ultrafast pulsed field-emission source with emission site selectivity. Simulations of local fields on the tip apex and of electron emission patterns based on the photo-field emission model describe our observations quantitatively.

%%%%%%%%%%%%%%%%%%%%%%%%%%%%%%%%%%%%%%%%%%%%%%%%%%%%%%%%%%%%%%%%%%%%%%%%%%%%%%%%%%%%%%%%%%%%%%%   fig1 schematic diagram of experiment     %%%%%%%%%%%%%%%%%%%%%%%%%%%%%%%%%%%%%%%%%%%%%%%%%%%%%%%%%%%%%%%%%%%%%%%%%%%%%%%%%%%%%%%%%%%%%%%%%%%%%%%%%%%%%%%%%%%%%
\begin{figure}[b]
\begin{center}
\includegraphics[scale=0.38]{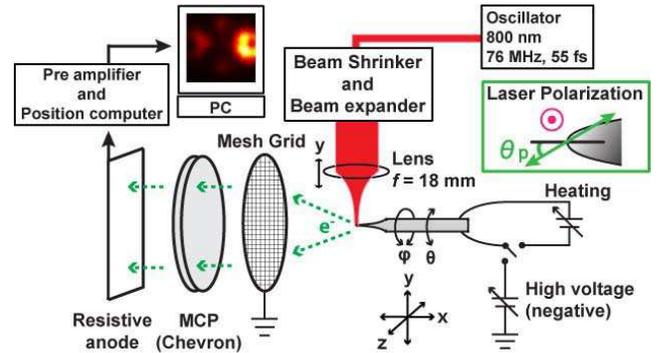}
\end{center}
\vskip -\lastskip \vskip -3pt
\caption{\label{fig:epsart}
Schematic diagram of the experimental setup. A tungsten tip is mounted inside a vacuum chamber. Laser pulses are generated outside the vaccum chamber. An aspherical lens is located just next to the tip to focus the laser onto the tip apex. Emitted electrons are detected by a position-sensitive detector in front of the tip. The polarization angle $\theta_{P}$ is defined in the inset (detailed explanations are given in the text).}
\label{fig:label-1}
\end{figure}
%%%%%%%%%%%%%%%%%%%%%%%%%%%%%%%%%%%%%%%%%%%%%%%%%%%%%%%%%%%%%%%%%%%%%%%%%%%%%%%%%%%%%%%%%%%%%%%%%%%%%%%%%%%%%%%%%%%%%%%%%%%%%%%%%%%%%%%%%%%%%%%%%%%%%%%%%%%%%%%%%%%%%%%%%%%%%%%%%%%%%%%%%%%%%%%%%%%%%%%%

%%%%%%%%%%%%%%%%%%%%%%%%%%%%%%%%%%%%%%%%%%%%%%%%%%%%%%%%%%%%%%%%%%%%%%%%%%%%%%%%%%%%%%%%%%%%% EXPERIMENTAL SETUP%%%%%%%%%%%%%%%%%%%%%%%%%%%%%%%%%%%%%%%%%%%%%%%%%%%%%%%%%%%%%%%%%%%%%%%%%%%%%%%%%%%%%%%%%%%%%%%%%%%%%%%%%%%%%%%%%%%%%%%%%%%
Fig. 1 schematically illustrates our experimental setup. A tungsten tip oriented towards the (011) direction is mounted inside a vacuum chamber ($3 \cdot 10^{-10}$ $mbar$). Laser pulses are generated in a Ti:sapphire oscillator (wave length: 800 $nm$, repetition rate: 76 $MHz$, pulse width: 55 $fs$) and introduced into the vacuum chamber.  The laser light was focused to 4 $\mu m$ ($1/e^2$ radius) onto the tip apex. Linearly polarized laser light was used. The polarization vector can be changed within the transversal (x, z) plane by using a $\lambda/2$ plate. As shown in the inset, where the laser propagates towards the reader's eye as denoted by the red mark, the polarization angle $\theta_{P}$ is defined by the angle between the tip axis and the polarization vector. The tip can be heated to clean the apex and also negatively biased for field emission. A position-sensitive detector in front of the tip measures the emission patterns. The spatial resolution of field emission microscopy (FEM) is approximately 3 $nm$~\cite{gomer93}. The tip holder can move along three linear axes (x, y, z) and has two rotational axes for azimuthal ($\varphi$, around the tip axis) and polar ($\theta$, around the z axis) angles. $\theta$ is set so that the tip axis is orthogonal to the laser propagation axis with a precision of $\pm$ 1 degree, and the laser propagates parallel to the horizontal y axis within an error of $\pm$ 1 degree. In these experiments, the base line of the rectangular detector is approximately 20$^\circ$ off from the horizontal (y axis) incidence direction, which means that the laser propagation axis is inclined by 20$^\circ$ from the horizontal line in the observed laser-induced FEM images (see dashed red arrow in Fig. 2a).

%%%%%%%%%%%%%%%%%%%%%%%%%%%%%%%%%%%%%%%%%%%%%%%%%%%%%%%%%%%%%%%%%%%%%%%%%%%%%%%%%%%%%%%%%%%%%%%   fig2 experimental results   %%%%%%%%%%%%%%%%%%%%%%%%%%%%%%%%%%%%%%%%%%%%%%%%%%%%%%%%%%%%%%%%%%%%%%%%%%%%%%%%%%%%%%%%%%%%%%%%%%%%%%%%%%%%%%%%%%%%%
\begin{figure}[b!]
\begin{center}
\includegraphics[scale=0.36]{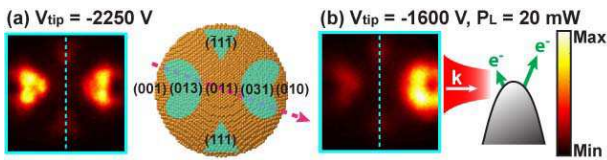}
\end{center}
\vskip -\lastskip \vskip -3pt
\caption{\label{fig:epsart}
Electron emission patterns without laser (a), and with laser irradiation (b). $V_{tip}$ indicates the DC potential applied to the tip and $P_{L}$ indicates the laser power measured outside the vacuum chamber. The inset of (a) shows the front view of the atomic structure of a tip apex based on a ball model. The inset of (b) shows a schematic side view of the laser-induced field emission geometry, in which green vectors indicate intensities of electron emission and the white arrow indicates the laser propagation direction. A dashed blue line denotes a mirror symmetry line of the atomic structure in each picture.}
\label{fig:label-2}
\end{figure}
%%%%%%%%%%%%%%%%%%%%%%%%%%%%%%%%%%%%%%%%%%%%%%%%%%%%%%%%%%%%%%%%%%%%%%%%%%%%%%%%%%%%%%%%%%%%%%%%%%%%%%%%%%%%%%%%%%%%%%%%%%%%%%%%%%%%%%%%%%%%%%%%%%%%%%%%%%%%%%%%%%%%%%%%%%%%%%%%%%%%%%%%%%%%%%%%%%%%%%

%%%%%%%%%%%%%%%%%%%%%%%%%%%%%%%%%%%%%%%%%%%%%%%%%%%%%%%%%%%%%%%%%%%%%%%%%%%%%%%%%%%%%%%%%%%%% EXPERIMENTAL RESULTS %%%%%%%%%%%%%%%%%%%%%%%%%%%%%%%%%%%%%%%%%%%%%%%%%%%%%%%%%%%%%%%%%%%%%%%%%%%%%%%%%%%%%%%%%%%%%%%%%%%%%%%%%%%%%%%%%%%%%%%%%%%
 The field emission pattern from the clean tungsten tip is shown in Fig. 2(a) where the most intense electron emission is observed around the (310)-type facets, and weaker emission from (111)-type facets. These regions are highlighted by green areas on the schematic front view of the tip apex in the inset of Fig. 2(a). The intensity map roughly represents a work function map of the tip apex: the lower the work function is, the more electrons are emitted. The relatively high work functions of (011)- and (001)-type facets \cite{kittel} suppress the field emission from those regions.

%%%%%%%%%%%%%%%%%%%%%%%%%%%%%%%%%%%%%%%%%%%%%%%%%%%%%%%%%%%%%%%%%%%%%%%%%%%%%%%%%%%%%%%%%%%%%%%   fig3 simulation results    %%%%%%%%%%%%%%%%%%%%%%%%%%%%%%%%%%%%%%%%%%%%%%%%%%%%%%%%%%%%%%%%%%%%%%%%%%%%%%%%%%%%%%%%%%%%%%%%%%%%%%%%%%%%%%%%%%%%%
\begin{figure*}[ht!]
\begin{center}
\includegraphics[scale=0.6]{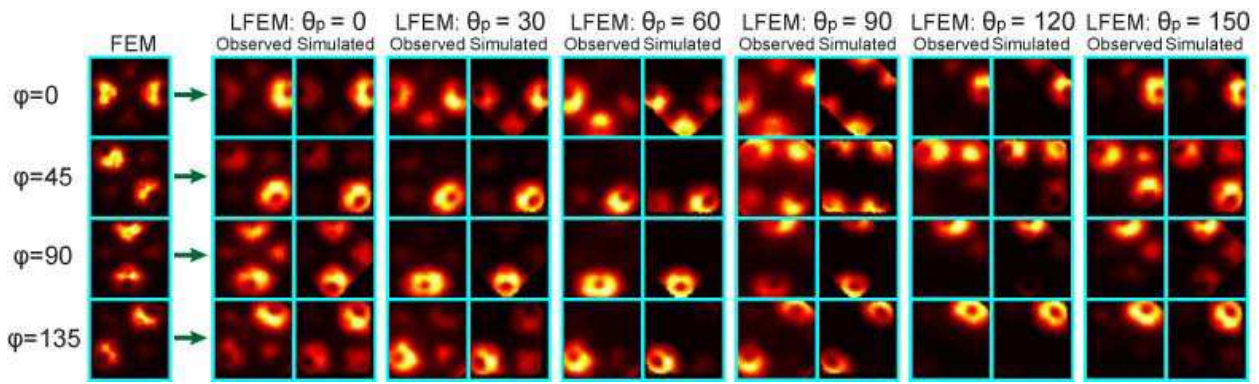}
\end{center}
\vskip -\lastskip \vskip -3pt
\caption{\label{fig:epsart}
Comparison of measured and simulated laser-induced FEM (LFEM) images for different light polarization angles $\theta_{P}$ and for different azimuthal orientations $\varphi$ of the tip. The leftmost column gives the FEM images without laser irradiation for four different azimuthal angles ($V_{tip}$ = -2250 $V$). For the same azimuthal angles, observed LFEM  images are shown as a function of polarization angle $\theta_{P}$ in 30$^\circ$ steps ($V_{tip} \approx$ -1500 $V$ and $P_{L}$ = 20 $mW$). The simulated LFEM images from the photo-field emission model, in which $V_{tip}$ and $P_{L}$ were set as in the corresponding experiments, are shown on the right-hand side of the observed LFEM images. The color scale and laser propagation direction are the same as in Fig. 2.}
\label{fig:label-3}
\end{figure*}
%%%%%%%%%%%%%%%%%%%%%%%%%%%%%%%%%%%%%%%%%%%%%%%%%%%%%%%%%%%%%%%%%%%%%%%%%%%%%%%%%%%%%%%%%%%%%%%%%%%%%%%%%%%%%%%%%%%%%%%%%%%%%%%%%%%%%%%%%%%%%%%%%%%%%%%%%%%%%%%%%%%%%%%%%%%%%%%%%%%%%%%%%%%%%%%%%%%%%%%%%%%%%

The laser-induced FEM (LFEM) image in Fig. 2(b), taken with the light polarization oriented parallel to the tip axis ($\theta_{P}$ = 0), shows a striking difference in symmetry compared to that of the FEM image in Fig. 2(a). Emission sites are the same in both cases, but the emission pattern becomes strongly asymmetric with respect to the shadow (right) and exposed (left) sides relative to the laser incidence direction. The most intense emission is
 observed on the shadow side as illustrated in the inset of Fig. 2(b). Actually, the laser pulses arrive at an oblique angle as indicated by the dashed red arrow in the inset of Fig. 2(a), which slightly affects the symmetry with respect to the central horizontal line in the observed LFEM images (see below).

We found experimentally a strong dependence of the electron emission patterns on the laser polarization direction and azimuthal orientations. Fig. 3 shows the LFEM patterns for different values of $\theta_{P}$ in 30$^\circ$ steps, and for four different azimuthal orientations $\varphi$ of the tip. The corresponding FEM images are also shown in the left most column. Throughout the whole image series, emission sites do not change, but intensities are strongly modulated resulting in highly asymmetric features. 

%%%%%%%%%%%%%%%%%%%%%%%%%%%%%%%%%%%%%%%%%%%%%%%%%%%%%%%%%%%%%%%%%%%%%%%%%%%%%%%%%%%%%%%%%%%%% Theoretical 1 %%%%%%%%%%%%%%%%%%%%%%%%%%%%%%%%%%%%%%%%%%%%%%%%%%%%%%%%%%%%%%%%%%%%%%%%%%%%%%%%%%%%%%%%%%%%%%%%%%%%%%%%%%%%%%%%%%%%%%%%%%%
When a laser pulse illuminates the metallic tip, surface EM waves~\cite{comment1} are excited on the laser-exposed side, which propagate around the tip apex along different paths as illustrated by red arrows in Fig. 4(a). As a result of interference between the excited waves, the electric fields show an asymmetric distribution over the tip apex, depending also on the laser polarization. To simulate the propagation of these surface EM waves and the resulting field distributions, we used the software package MaX-1 for solving Maxwell equations based on the Multiple Multipole Program~\cite{max1}. A droplet-like shape was employed as a model tip as shown in Fig. 4(b), with a radius of curvature of the tip apex of 100 $nm$, which is a typical value for a clean tungsten tip. The dielectric function $\epsilon$ of tungsten at 800 $nm$ was used: a real part $\Re(\epsilon) = 5.2$ and an imaginary part $\Im(\epsilon) = 19.4$ \cite{tungstenepsilon}. By using different droplet sizes it was verified that the model tip is long enough so as to mimick the infinite length of the real tip.

%%%%%%%%%%%%%%%%%%%%%%%%%%%%%%%%%%%%%%%%%%%%%%%%%%%%%%%%%%%%%%%%%%%%%%%%%%%%%%%%%%%%%%%%%%%%%%%   fig4 polarization data    %%%%%%%%%%%%%%%%%%%%%%%%%%%%%%%%%%%%%%%%%%%%%%%%%%%%%%%%%%%%%%%%%%%%%%%%%%%%%%%%%%%%%%%%%%%%%%%%%%%%%%%%%%%%%%%%%%%%%
\begin{figure}[bt]
\begin{center}
\includegraphics[scale=0.38]{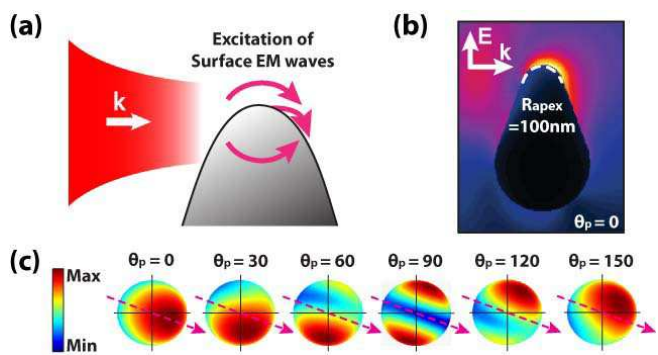}
\end{center}
\vskip -\lastskip \vskip -3pt
\caption{\label{fig:epsart}
Schematic illustration of the excitation and interference of surface EM waves (a). The calculated time-averaged field distribution around the model tip is shown in a linear color scale in (b) for $\theta_{P}=0^\circ$. Highest field values are represented in yellow. In (c) the time-averaged field distributions are given in a front view of the model tip for different polarization directions $\theta_{P}$. The laser propagation direction is indicated by red arrow, it is the same as in the experiment.}
\label{fig:label-4}
\end{figure}

%%%%%%%%%%%%%%%%%%%%%%%%%%%%%%%%%%%%%%%%%%%%%%%%%%%%%%%%%%%%%%%%%%%%%%%%%%%%%%%%%%%%%%%%%%%%%%%%%%%%%%%%%%%%%%%%%%%%%%%%%%%%%%%%%%%%%%%%%%%%%%%%%%%%%%%%%%%%%%%%%%%%%%%%%%%%%%%%%%%%%%%%%%%%%%%%%%%%%%%%

Fig. 4(b) shows the calculated time-averaged field distribution over a cross section of the model tip, where the polarization vector is parallel to the tip axis ($\theta_{P}=0^\circ$). The laser propagates from left to right, and was focused at the tip apex. The field distribution is clearly asymmetric with respect to the tip axis, with a maximum on the shadow side of the tip. This is consistent with our observations in Fig. 2(b) where the field emission is enhanced on the shadow side. Fig. 4(c) shows, in a front view, time-averaged field distribution maps from the white dashed line region of the model tip in Fig. 4(b). This area corresponds roughly to the observed area in our experiments. The field distribution changes strongly depending on the polarization angle.

%%%%%%%%%%%%%%%%%%%%%%%%%%%%%%%%%%%%%%%%%%%%%%%%%%%%%%%%%%%%%%%%%%%%%%%%%%%%%%%%%%%%%%%%%%%%% Theoretical 2 %%%%%%%%%%%%%%%%%%%%%%%%%%%%%%%%%%%%%%%%%%%%%%%%%%%%%%%%%%%%%%%%%%%%%%%%%%%%%%%%%%%%%%%%%%%%%%%%%%%%%%%%%%%%%%%%%%%%%%%%%%%
From the calculated local fields, we futher simulated the LFEM images by considering the photo-field emission processes. The current density $j_{calc}$ of field emission can be described in the Fowler-Nordheim theory based on the free-electron model as follows~\cite{gomer93,murphy56,young59,fursey03},
%%%%%%%%%%%%%%%%%%%%%%%%%%%%%%%%%%%%%%%%%%%%%%%%%%%%%%%%%%%%%%%%%%%%%%%%%%%%%%%%%%%%%%%%%%%%%%%   phonon intensity formula     %%%%%%%%%%%%%%%%%%%%%%%%%%%%%%%%%%%%%%%%%%%%%%%%%%%%%%%%%%%%%%%%%%%%%%%%%%%%%%%%%%%%%%%%%%%%%%%%%%%%%%%%%%%%%%%%%%%%%
\begin{eqnarray}
j_{calc} =  \frac{em}{2 \pi^{2} \hbar^{3}} \int_{-\infty}^{\infty} \int_{-W_a}^{W=E} D(W, \Phi, F)f(E)\,d W d E.
\end{eqnarray}
%%%%%%%%%%%%%%%%%%%%%%%%%%%%%%%%%%%%%%%%%%%%%%%%%%%%%%%%%%%%%%%%%%%%%%%%%%%%%%%%%%%%%%%%%%%%%%%%%%%%%%%%%%%%%%%%%%%%%%%%%%%%%%%%%%%%%%%%%%%%%%%%%%%%%%%%%%%%%%%%%%%%%%%%%%%%%%%%
Here, $e$ is the electron charge and $m$ the electron mass, $W_{a}$ is the effective constant potential energy inside the metal, W is the normal energy with respect to the surface, and E is the total energy. Important factors are $D(W, \Phi, F)$ and $f(E)$. $D(W, \Phi, F)$ is the probability that an electron with the normal energy W penetrates the surface barrier. It depends exponetially on the triangular-shaped potential barrier above W, which is determined by the work function $\Phi$ and the applied electric field $F$. $f(E)$ is an electron distribution function which is the Fermi-Dirac distribution for an equilibrium state. 

In the photo-field emission model, the Fermi-Dirac distribution is strongly modified by the electron-hole pair excitations due to multiphoton absorption. Here the electron distribution becomes a nonequilibrium distribution, characterized by a steplike profile with steps separated in energy by the photon energy $h\nu$ = 1.55 $eV$, thus extending far above the Fermi level $E_{F}$~\cite{wu08, lisowski04}. Each step height in the occupation numbers depends on the number of photons $N$ available per area during a pulse, given by $N = \frac{I} {h \nu f_{rep}}$ where $I$ is the laser intensity and $f_{rep}$ is the repetition rate of laser pulses. In the vicinity of the tip we have $I\propto F_{laser}^2$ where $F_{laser}$ is the enhanced electric field on the tip apex as illustrated in Fig. 4(c) \cite{merschdorf04}. For example, one-photon absorption creates a step from $E_{F}$ to $E_{F}+h\nu$ by exciting elecrons from the region $E_{F}-h\nu$ to $E_{F}$. The step height is equal to $\alpha N$ where $\alpha$ is a mean photoelectric cross section. Absorption of a second photon creates a step from $E_{F}+h\nu$ to $E_{F}+2h\nu$ by exciting elecrons from $E_{F}$ to $E_{F}+h\nu$, with a step height equal to $(\alpha N)^{2}$. We included absorption of up to four photons.

There are three adjustable parameters in our calculations of $j_{calc}$: $\Phi$, $F$, and $\alpha N$. They are all functions of position on the tip apex. $\Phi$ and $F$ maps on the tip apex were obtained from the measured FEM images, which were symmetrized to have the ideal two-fold symmetry, representing the current density $j_{exp}$ as a function of position on the tip apex. Starting from a relative $F$ distribution obtained by MaX-1, we obtained the $\Phi$ map by inserting $F$ into Eq. (1) and postulating $j_{exp} -  j_{calc} = 0$. The resulting $\Phi$ map then was compared to known values for several surface facets and adjusted by scaling the absolute values of $F$ up or down accordingly by multiplying $F$ distribution with a constant factor. Thus, a full $\Phi$ map and absolute values for $F$ were obtained. The resulting $\Phi$ values are approximately 4.9 $eV$, 4.6 $eV$ and 4.45 $eV$ for (011), (001) and (310) surfaces, respectively, which are in fair agreement with known values \cite{kittel, Mendenhall34}. A field strength $F$ of $2.25$ $V/nm$ results at the tip apex for the FEM image taken with $V_{tip}=-2250$ $V$, which is a typical value for FEM. The LFEM experiments were carried out with a reduced tip voltage $V_{tip} \approx -1500 V$, hence we used a corresponding value of 1.5 $V/nm$ in the LFEM simulations.

Substituting the obtained $\Phi$ and $F$ distribution maps into Eq. (1), and using a nonequilibrium electron distribution $f(E)$, the absolute values of $\alpha N$ over the tip apex was determined by fitting the measured total current from the right-hand side (310) facet of the LFEM image in Fig. 2(b). The resulting maximum value for $\alpha N$ was $2.5 \cdot 10^{-6}$. By substituting all the adjusted parameters into Eq. (1), we could simulate all the LFEM images. The calculated current densities on the tip apex were projected to the flat screen by following the static field lines. The simulated images can now be directly compared to the experimental images (Fig. 3): they are in excellent agreement in every detail. This comparison clearly demonstrates that the observed strongly asymmetric features originate from the modulation of the local photo-fields.

%%%%%%%%%%%%%%%%%%%%%%%%%%%%%%%%%%%%%%%%%%%%%%%%%%%%%%%%%%%%%%%%%%%%%%%%%%%%%%%%%%%%%%%%%%%%%%%   fig5 line profile   %%%%%%%%%%%%%%%%%%%%%%%%%%%%%%%%%%%%%%%%%%%%%%%%%%%%%%%%%%%%%%%%%%%%%%%%%%%%%%%%%%%%%%%%%%%%%%%%%%%%%%%%%%%%%%%%%%%%%
\begin{figure}[t]
\begin{center}
\includegraphics[scale=0.35]{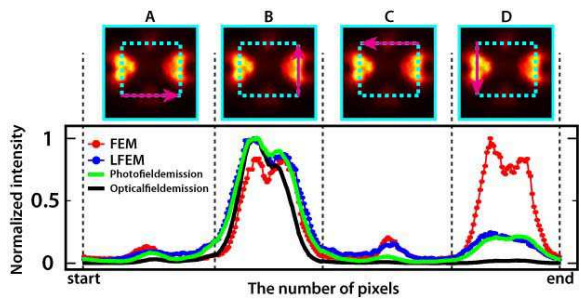}
\end{center}
\vskip -\lastskip \vskip -3pt
\caption{\label{fig:epsart}
Line profiles extracted from the observed FEM images (red line with dots), LFEM images (blue line with dots), and from LFEM images simulated by the photo-field emission model (green line) and the optical field emission model (black line) at [$\varphi=0^\circ$, $\theta_{P}=0^\circ$]. The whole scanned line corresponds to the unfolded rectangle indicated by the dashed blue line in the FEM figures above, and the corresponding sides are indicated by red arrows. Each line profile has been normalized by the maximum value.}
\label{fig:label-5}
\end{figure}
%%%%%%%%%%%%%%%%%%%%%%%%%%%%%%%%%%%%%%%%%%%%%%%%%%%%%%%%%%%%%%%%%%%%%%%%%%%%%%%%%%%%%%%%%%%%%%%%%%%%%%%%%%%%%%%%%%%%%%%%%%%%%%%%%%%%%%%%%%%%%%%%%%%%%%%%%%%%%%%%%%%%%%%%%%%%%%%%%%%%%%%%%%%%%%%%%%%%%%%%

Finally, we simulated the LFEM images also for the optical field emission process and compared the resulting intensity distributions with those of the photo-field emission model. In the optical field emission model, the Fermi-Dirac distribution is not modified, but instead the electric field $F$ in Eq. (1) is expressed as $F=F_{DC}+F_{laser}^{\perp}$ where $F_{DC}$ is the DC electric field and $F_{laser}^{\perp}$ is the local photo-field component normal to the local tip surface. The absolute values for $F_{laser}^{\perp}$ on the tip apex were determined in the same way as described above for the $\alpha N$ values. The resulting maximum normal field component was 0.68 $V/nm$. 

Fig. 5 shows line profiles extracted from the observed FEM and LFEM images, and from the corresponding simulations for both photo-field emission and optical field emission models at [$\varphi=0^\circ$, $\theta_{P}=0^\circ$], which were all normalized by their maximum value. The whole scanned line corresponds to the unfolded rectangle indicated in the example FEM image; it provides sections through all major intensity features of these data. The measured LFEM profile clearly shows the asymmetric feature as seen in the regions B and D. The photo-field emission model (green line) catches this asymmetry much more quantitatively than the optical field emission model (black line), as can be best seen in region D. Therefore, the local fields in our experiment are still weak enough such that the photo-field emission process is the dominant one.

%%%%%%%%%%%%%%%%%%%%%%%%%%%%%%%%%%%%%%%%%%%%%%%%%%%%%%%%%%%%%%%%%%%%%%%%%%%%%%%%%%%%%%%%%%% SUMMARY %%%%%%%%%%%%%%%%%%%%%%%%%%%%%%%%%%%%%%%%%%%%%%%%%%%%%%%%%%%%%%%%%%%%%%%%%%%%%%%%%%%%%%%%%%%%%%%%%%%%%%%%%%%%%%%%%%%%%%%%%%%%%%%%%%%%%%%%%%%%%%%%%%%%%%%%%%%
In summary, we have demonstrated the realization of an ultrafast pulsed field-emission source with convenient control of nanometer sized emission sites by the laser polarization and incident laser angle relative to the azimuthal orientations. The photo-field emission model well reproduced our observations. Maybe the most interesting applications will arise when two laser pulses with different polarizations or paths are used. Electron emission from two independent sources but originating from the same quantum state could be studied \cite{oshima02,cho04}. This should create new opportunities for addressing fundamental questions in quantum mechanics such as anticorrelation of electron waves in vacuum \cite{klesel02}, or for new directions in electron holography \cite{tonomura87}.

%%%%%%%%%%%%%%%%%%%%%%%%%%%%%%%%%%%%%%%%%%%%%%%%%%%%%%%%%%%%%%%%%%%%%%%%%%%%%%%%%%%%%%%%%%%%%              ACKNOWLEDGEMENT                %%%%%%%%%%%%%%%%%%%%%%%%%%%%%%%%%%%%%%%%%%%%%%%%%%%%%%%%%%%%%%%%%%%%%%%%%%%%%%%%%%%%%%%%%%%%%%%%%%%%%%%%%%
We acknowledge many useful discussions with Prof. H. W. Fink, Dr. T. Ishikawa, and Dr. K. Kamide. This work was supported in part by the Japan Society for the Promotion of Science (JSPS), and the Swiss National Science Foundation (SNSF).

%%%%%%%%%%%%%%%%%%%%%%%%%%%%%%%%%%%%%%%%%%%%%%%%%%%%%%%%%%%%%%%%%%%%%%%%%%%%%%%%%%%%%%%%%%%%%              REFERENCES                %%%%%%%%%%%%%%%%%%%%%%%%%%%%%%%%%%%%%%%%%%%%%%%%%%%%%%%%%%%%%%%%%%%%%%%%%%%%%%%%%%%%%%%%%%%%%%%%%%%%%%%%%%
%\include{HiroReferences}

\end{document}